\documentclass[showpacs,preprintnumbers,prb,amsmath,amssymb,superscriptaddress,twocolumn]{revtex4}
\usepackage{endnotes}
\usepackage{dcolumn}
\usepackage{bm}
\usepackage{graphicx}
\usepackage{color}
\usepackage{amssymb}

\begin{document}

\preprint{APS/123-QED}

\title{Spectroscopic study of metallic magnetism in Nb$_{1-y}$Fe$_{2+y}$} 

\author{D. Rauch}
\author{M. Kraken}
\author{F. J. Litterst}
\author{S. S\"ullow}
\affiliation{Institute of Condensed Matter Physics, Technische Universit\"at Braunschweig, D-38106 Braunschweig, Germany}

\author{H. Luetkens}
\affiliation{Laboratory for Muon Spin Spectroscopy, Paul Scherrer Institute, CH-5232 Villigen, Switzerland}

\author{M. Brando}
\author{T. F\"orster}
\author{J. Sichelschmidt}
\affiliation{Max-Planck-Institute for Chemical Physics of Solids, D-01187 Dresden, Germany}

\author{A. Neubauer} 
\author {C. Pfleiderer}
\affiliation{Physik Department E21, Technische Universit\"at M\"unchen,
 D-85748 Garching, Germany}

\author {W. J. Duncan}
\affiliation{Department of Physics, Royal Holloway, University of London, Egham TW20 0EX, United Kingdom}

\author{F. M. Grosche}
\affiliation{Cavendish Laboratory, University of Cambridge, Cambridge CB3 0HE, United Kingdom}
\date{\today}

\begin{abstract}
We have investigated single crystals and polycrystals from the series Nb$_{1-y}$Fe$_{2+y}$, $-0.004 \leq y \leq 0.018$ by electron spin resonance, muon spin relaxation and M\"ossbauer spectroscopy. Our data establish that at lowest temperatures all samples exhibit bulk magnetic order. Slight Fe-excess induces low-moment ferromagnetism, consistent with bulk magnetometry, while Nb--rich and stoichiometric NbFe$_2$ display spin density wave order with small magnetic moment amplitudes of the order $\sim 0.001 - 0.01 \mu_B/{\rm Fe}$. This provides microscopic evidence for a modulated magnetic state on the border of ferromagnetism in NbFe$_2$.
\end{abstract}


\maketitle

\section{Introduction}

Ferromagnetic quantum phase transitions in metals represent a topic of current research interest both from the theoretical and experimental side, this especially under the aspect of non-Fermi liquid behavior in transition metal compounds \cite{Pfleiderer94,Grosche95}. The underlying question whether a ferromagnetic quantum critical point can exist in clean band magnets remains controversial so far. Based on fundamental considerations \cite{belitz05, chubukov04, conduit09, kirkpatrick12}, it was argued that ferromagnets adopt one of two possible scenarios on approaching the putative quantum critical point: either the transition into the ferromagnetic state becomes discontinuous (first order), or the low temperature state adopts a modulation (ordering wave vector $Q \neq 0$), causing some form of spin density wave order (SDW).

An interesting candidate for the SDW scenario is NbFe$_2$ \cite{brando08}. In the Nb$_{1-y}$Fe$_{2+y}$ composition series, the magnetic state depends sensitively on composition \cite{shiga87,yamada88,crook95,moroni09}. Nb--rich ($y < -0.02$) and Fe--rich ($y > 0.005$) Nb$_{1-y}$Fe$_{2+y}$ are weakly ferromagnetic (FM) at low temperature, which may be attributed to changes in the electronic structure \cite{Subedi10,Tompsett10,Neal11,Alam11}. At intermediate compositions $y \simeq 0$ in Nb$_{1-y}$Fe$_{2+y}$, however, samples are not FM, although they still appear to order magnetically as shown by susceptibility measurements, and which was tentatively interpreted as a SDW state. Signatures of Fermi liquid breakdown have been reported near the critical concentration $y = -0.015$, where the ordering temperature of the putative SDW state extrapolates to zero \cite{brando08}. 

Despite many experimental efforts, microscopic evidence for the existence of a broken symmetry magnetic low temperature state in stoichiometric NbFe$_2$ has so far been lacking. NMR studies on polycrystals \cite{yamada88} require magnetic fields exceeding the critical field of the ordered phase for some grain orientations, whereas the small size of the ordered moment, estimated to less than $0.1 \mu_B/{\rm Fe}$ from magnetization measurements in the FM state, imply a tiny signal size in neutron scattering. Here, we present low temperature electron spin resonance (ESR), muon spin relaxation ($\mu$SR) and M\"ossbauer spectroscopy experiments on single and polycrystalline Nb$_{1-y}$Fe$_{2+y}$ in the alloying range $-0.004 \leq y \leq 0.018$, thus covering the range from SDW to FM ground state. 

Our study establishes the bulk nature and spatial homogeneity of these phases at the lowest temperatures. We determine the temperature evolution and magnitude of the internal magnetic field, and assess the size of the ordered magnetic moments in the different phases. Our findings confirm the FM nature of the low temperature state in Fe-rich NbFe$_2$. Further, they demonstrate the existence of a modulated magnetic phase (SDW) at low temperature in near-stoichiometric NbFe$_2$ and in an intermediate temperature range above the FM state in Fe-rich NbFe$_2$. Moreover, they indicate that the ordered moment of the modulated magnetic phase diminishes more rapidly as a function of Nb concentration than the associated ordering temperature on approaching the quantum critical point.

\section{Experimental details}

Single crystals Nb$_{1-y}$Fe$_{2+y}$ were grown in a UHV compatible mirror furnace from polycrystalline precursor rods prepared by radio-frequency induction melting, and characterized by Laue diffractometry, transport measurements and magnetometry \cite{PhDNeubauer, PhDDuncan, Duncan10}. The samples used in this study correspond closely to similar single crystals from the same growth runs, which were characterized extensively by bulk experiments \cite{Duncan10,Friedemann13}. They include Fe-rich Nb$_{0.982}$Fe$_{2.018}$ (with a SDW phase below T$_N \sim 36$\,K and a FM transition T$_C \sim 32$\,K (determined from magnetization measurements; Fig.~\ref{fig:asymmetryTF}b), stoichiometric NbFe$_{2}$ (T$_N \sim 14$\,K) and Nb--rich Nb$_{1.004}$Fe$_{1.996}$ (T$_N \sim 8$\,K). An iron-rich polycrystal Nb$_{0.984}$Fe$_{2.016}$ (T$_N \sim 30$\,K, T$_C \sim 22$\,K) was used for M\"ossbauer spectroscopy.

ESR experiments were carried out with a standard continuous-wave spectrometer between 5 and 300\,K. We measured the power $P$ absorbed by the sample from a transverse magnetic microwave field ($\nu \approx 9.4$\,GHz) as a function of an external, static magnetic field $B$. A lock-in technique (with a modulated field at 100\,kHz) was used to improve the signal-to-noise ratio, which yields the derivative of the resonance signal $dP/dB$. $\mu$SR experiments in zero field (ZF) and in a weak transverse applied field (wTF) (external field of 5\,m$T$ applied at a $90^\circ$ angle relative to the polarized muon spin) have been performed between 1.8 and 140\,K using the GPS facility of the Swiss Muon Source at the Paul Scherrer Institut, Villigen. M\"{o}ssbauer spectroscopy has been performed between $4$ and $90$\,K using a conventional M\"{o}ssbauer set--up.

\section{Results: ESR}

\begin{figure}[t]
\centering
\includegraphics[width=1\columnwidth]{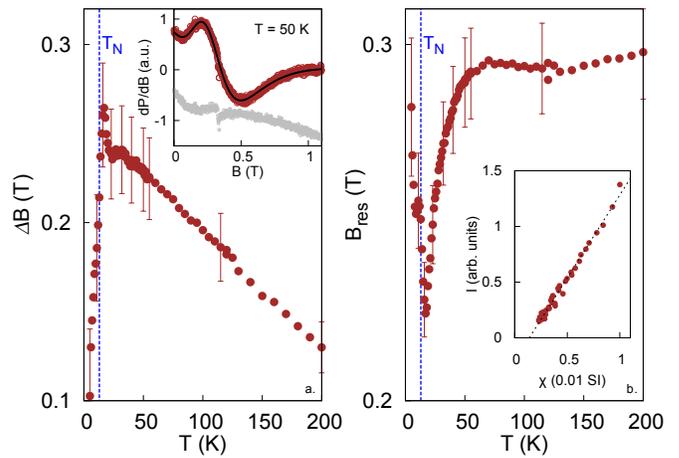}
  \caption{(Color online) (a) Paramagnetic resonance signal (inset) at T $= 50$\,K of NbFe$_{2}$ and temperature dependent linewidth $\Delta B({\rm T})$. Solid line describes the spectrum by a Lorentzian shape plus a weak background (gray spectrum) responsible for the initial drop of the signal. (b) Temperature dependence of the resonance field $B_{\rm res}({\rm T})$ perpendicular to the magnetic easy axis (less than the critical field $\simeq 3$\,$T$ required to suppress T$_N$ to zero). Inset: Integrated ESR intensity vs. ac-susceptibility taken at 0.25\,$T$ between 20 and 100\,K.} 
\label{figure_ESR}
\end{figure}

The ESR measurements on a stoichiometric single crystal of NbFe$_2$ show well-defined resonances probed in a skin depth of about $2\, \rm \mu m$ (at 20\,K) at a crystal size of $1 \times 0.9 \times 0.4$\,mm$^3$. The inset of Fig.~\ref{figure_ESR}a. shows a typical spectrum in the paramagnetic regime together with a single line metallic Lorentzian shape for a resonance with $g=2.1$. Such resonance spectra taken on the single crystal NbFe$_2$ were fitted with a metallic Lorentzian shape \cite{Joshi04} with the parameters linewidth $\Delta B$ (half width at half maximum of the microwave power absorbed by ESR) and resonance field $B_{res}$ of an ESR line with a Lorentzian shape:
\begin{eqnarray}
\frac{dP(B)}{dB} = Amp \cdot \left\{ \frac{\alpha(1-x^2) - 2 \cdot x }{(1+x^2)^2} + \frac{-\alpha(1-y^2) - 2 \cdot y }{(1+y^2)^2}\right\} \label{eq:ESR}
\end{eqnarray}
Here, $Amp$ is the amplitude, $\alpha$ is the asymmetry parameter ($D/A$, describing the microwave dispersion relative to absorption in metallic samples). Further, we use $x = \frac{B-B_{res}}{\Delta B}$ and $y = \frac{B+B_{res}}{\Delta B}$. The second term describes the influence of the counter rotating component of the linearly polarized microwave field, which is relevant in the case of $\Delta B \gtrsim B_{res}$, as we will demonstrate below. We determined the ESR intensity by integrating the ESR spectra and dividing by the temperature dependent skin depth.

In general, a conduction electron spin resonance is described by a Dysonian lineshape which takes into account the spin diffusion, penetration depth and sample thickness \cite{dyson55a,feher55}. In the limit of a much faster spin relaxation than the spin diffusion through the skin depth (as it is realized in NbFe$_{2}$) a regular local moment ESR lineshape is expected. In this case, however, it is important to note that microwave dispersion effects occur due to the skin effect. Hence, an asymmetric line, characterized by the ratio of dispersion to absorption, $D/A$, is found, {\it viz.}, a ''metallic'' Lorentzian.   

\begin{figure}[t]
 \centering
 \includegraphics[width=0.95\columnwidth]{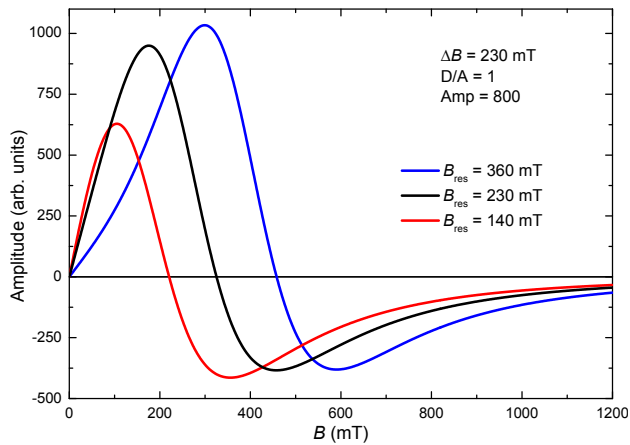}
  \caption{(Color online) ESR lineshape variation for various resonance fields $B_{\rm res}$ being comparable to the linewidth $\Delta B$. The lineshape is a metallic Lorentzian with D/A=1 reflecting the case of a small microwave penetration depth compared to the sample thickness and a slow spin diffusion.}
 \label{fig1sup}
\end{figure}

Another crucial effect to the lineshape arises if the ESR linewidth is very broad and of the order of the resonance field. Then the ESR lineshape may considerably deviate from the shape expected for narrow lines because the counter rotating component of the linearly polarized microwave field becomes relevant \cite{Duncan10}. Fig.~\ref{fig1sup} illustrates this situation for an ESR line with parameters corresponding to those of NbFe$_{2}$. Note the change in the portions of positive and negative contributions to the lineshape and compare to the case of NbFe$_{2}$ shown in the inset of Fig.~\ref{figure_ESR}a.  

Linewidth $\Delta B$ and resonance field $B_{\rm res}$ as determined from fits of the ESR spectra to Eq.~\ref{eq:ESR} show very similar temperature dependencies compared to those reported previously in polycrystalline samples \cite{foerster10}. The linearity between the ESR intensity (being a microscopic probe of the spin susceptibility in the skin depth and the magnetic susceptibility (see inset of Fig.~\ref{figure_ESR}b) indicates that the observed resonance reflects bulk magnetic properties. 

\begin{figure}[t]
 \centering
 \includegraphics[width=0.9\columnwidth]{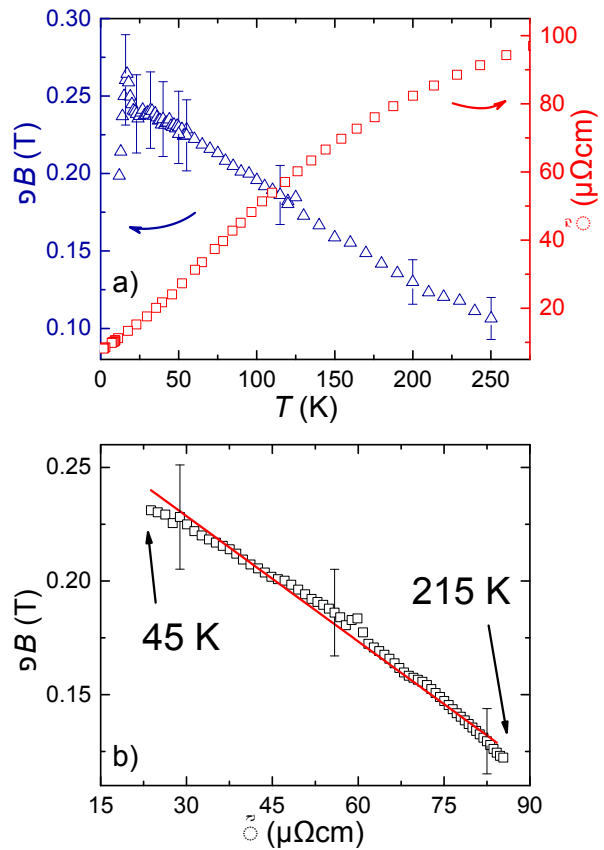}
 \caption{(Color online) The upper panel shows the temperature dependence of the ESR linewidth $\Delta B$ and the electric resistivity~$\rho$. The lower panel displays $\Delta B$ as function of $\rho$, with temperature as implicit parameter. The solid line suggests a linearity between $\Delta B$ and $\rho$ in a wide temperature range.}
 \label{fig2sup}
\end{figure}

The clearest evidence for the bulk origin of the signal is that $\Delta B$ decreases (almost linearly) with increasing temperature following the opposite behavior of the electrical resistivity $\rho$ in the same range of temperatures, which namely increases with increasing T \cite{Friedemann13,moess14}. This is explicitely demonstrated in Fig.~\ref{fig2sup}, where we combine both quantities, $\Delta B$ and $\rho$, in the same plot. A direct relation between $\Delta B$ and $\rho$ is a characteristic hallmark of a conduction electron spin resonance (CESR) and would be completely unexpected for impurity spins or coupled spins of impurity and itinerant electrons. 

More specifically, correlated band magnets like TiBe$_2$ are examples for a CESR which can be analyzed according to the Elliot-Yafet theory \cite{shaltiel88a}. A generalized version of this theory \cite{simon08,dora09a} takes the particular electronic band structure into account and allows to understand the relation between linewidth $\Delta B$ and resistivity $\rho$. We considered the special preconditions of the electronic band structure of NbFe$_2$ \cite{Subedi10,Tompsett10,Neal11,Alam11} and found for $\Delta B$ a behavior almost linear to $\rho$, but with opposite prefactors. To illustrate this we show the resistivity data of NbFe$_2$ from Ref.~\cite{Friedemann13} in the combined plot with the ESR linewidth vs. temperature, see Fig.~\ref{fig2sup}a, and with temperature as implicit parameter, see Fig.~\ref{fig2sup}b. A similar behavior was as also found for MgB$_2$ \cite{moess14} and the correlated narrow band metal Rb$_3$C$_{60}$ \cite{dora09a} in a wide temperature range. Indeed, for NbFe$_2$ a prerequisite of the generalized Elliot-Yafet theory is fulfilled, namely that the quasiparticle scattering rate is comparable to the energy separation of two neighboring spin-orbit split bands close to the Fermi level \cite{moess14}. Taking these observations as strong indications for a bulk CESR in NbFe$_2$ we continue with discussing the effects of the ground state phase.

Towards low temperatures and crossing the ordering temperature T$_N \sim 12$\,K, $\Delta B({\rm T})$ and $B_{\rm res}({\rm T})$ show strong changes which suggest a drastic change in internal fields at T$_N$ and which indicate the presence of a low temperature magnetically ordered state. Such strong changes in $B_{\rm res}({\rm T})$ may be expected in particular for systems with strong magnetic anisotropies \cite{huber} like NbFe$_2$ \cite{kurisu}.

\section{Results: $\mu$SR}

\begin{figure}[t]
	\centering
		\includegraphics[width=1\linewidth]{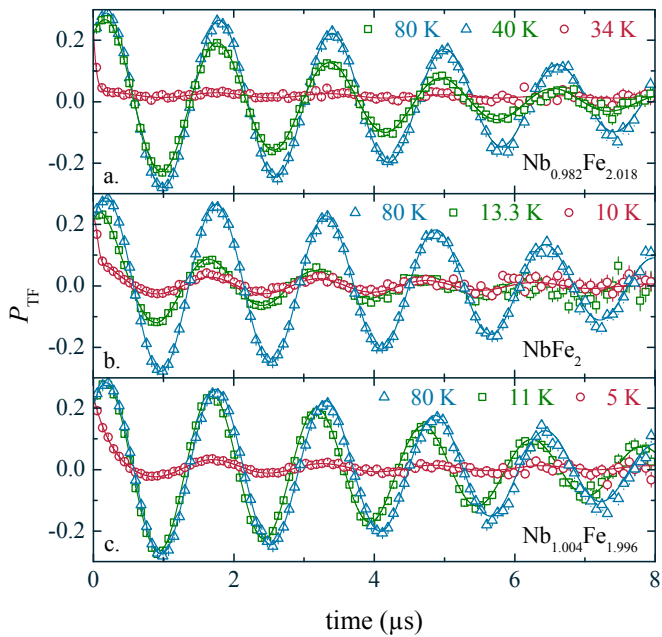}
	\caption{(Color online) Weak transverse field $\mu$SR asymmetry spectra between 5 and 80\,K in an external field of 5\,m$T$ for (a) Nb$_{0.982}$Fe$_{2.018}$, (b) NbFe$_2$, and (c) Nb$_{1.004}$Fe$_{1.996}$. Solid lines are fits to the data. Nb$_{0.982}$Fe$_{2.018}$ has been measured in an applied field of 3\,m$T$ for T $\geq 40$\,K. For all samples a small sample holder contribution (10 $\%$ of the total signal) has been included in the fit.}
	\label{fig:TF}
\end{figure}

The $\mu$SR data provide comprehensive microscopic information about the low temperature magnetic states in Nb$_{1-y}$Fe$_{2+y}$. Fig.~\ref{fig:TF} displays the wTF spectra of Nb$_{1-y}$Fe$_{2+y}$ between 5 and 80\,K. At high temperatures, the muons precess with frequency $\omega$ in the external field, giving rise to an oscillatory asymmetry signal $P_{\mathrm{TF}}$  
\begin{eqnarray}
	P_{\mathrm{TF}} = a \cos \left( \omega t + \phi \right) e^{-\frac{1}{2} \left( \sigma t \right)^2} e^{-\lambda_{\mathrm{TF}} t}. 	\label{eq:TFh}
\end{eqnarray}
Eq.~\ref{eq:TFh} implies a local field distribution caused by nuclear dipoles and an additionally operative electronic relaxation process. Here, $a$ denotes the asymmetry parameter, $\phi$ the phase shift, $\sigma$ the width of a local magnetic field distribution and $\lambda_{\mathrm{TF}}$ the damping rate. 

\begin{figure}[t]
	\centering
		\includegraphics[width=1\linewidth]{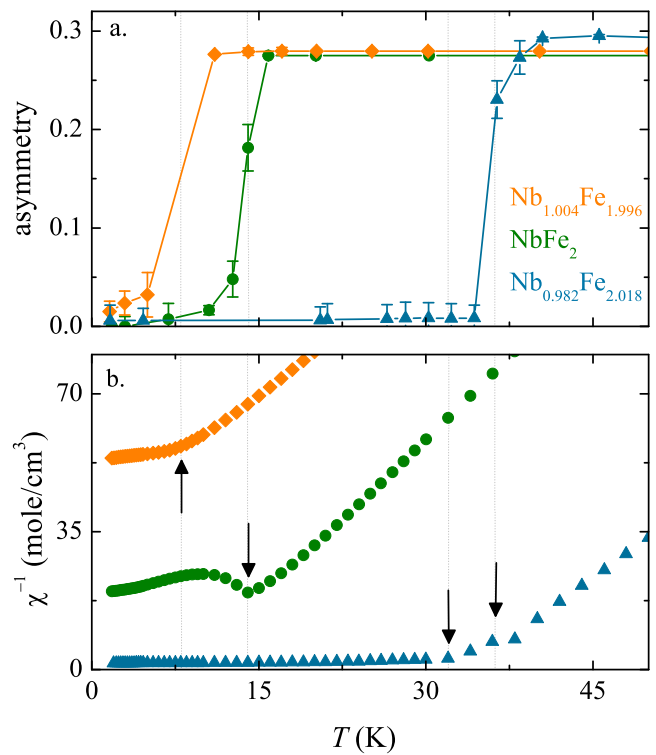}
	\caption{(Color online) Temperature dependence of the asymmetry parameter from wTF measurements (a) and inverse susceptibility $\chi^{-1}$ (b) in Nb$_{1-y}$Fe$_{2+y}$, $y=+0.018$ (triangle), $0.0$ (circle) and $-0.004$ (diamond).}
	\label{fig:asymmetryTF}
\end{figure}

At low temperature, the signals are strongly damped and acquire an additional precession term from local internal fields (see Eqs.~\ref{eq:iFld} and \ref{eq:iBsl}). From fits of the data we obtain the temperature dependence of the wTF asymmetry parameter, which is proportional to the paramagnetic volume fraction (Fig.~\ref{fig:asymmetryTF}a). The complete loss of asymmetry for all samples at the lowest temperatures proves the bulk nature of the magnetically ordered phases. The transition temperatures are determined as 36\,K in Nb$_{0.982}$Fe$_{2.018}$, 14\,K in NbFe$_{2}$ and 8\,K in Nb$_{1.004}$Fe$_{1.996}$, in good agreement with  bulk studies \cite{brando08, moroni09, PhDNeubauer, PhDDuncan, Duncan10, Friedemann13}.

\begin{figure}[t]
	\centering
		\includegraphics[width=1\linewidth]{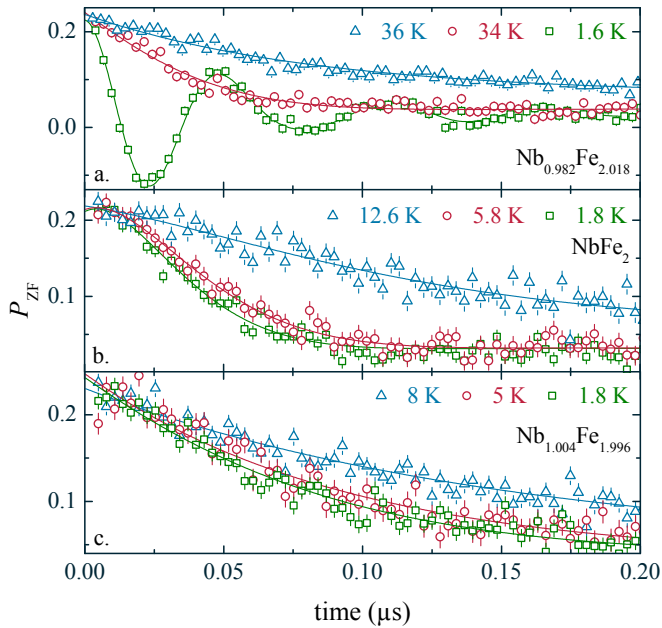}
	\caption{(Color online) Zero field $\mu$SR asymmetry spectrum between 1.8 and 36\,K of Nb$_{1-y}$Fe$_{2+y}$, $y=+0.018$ (a), $0.0$ (b) and $-0.004$ (c). Solid lines are fits according to Eqs.~\ref{eq:iFld} and \ref{eq:iBsl}.}
	\label{fig:ZFlowT}
\end{figure}

More detailed information is obtained from zero field (ZF) $\mu$SR experiments (Fig.~\ref{fig:ZFlowT}). Above the critical temperatures, the data for all samples can be fitted by a dynamic Gaussian Kubo--Toyabe--type function \cite{moess14,Hayano79}. Inside the magnetically ordered phases, the relaxation behavior depends on the nature of the ordered state. 

\begin{figure}[t]
	\centering
		\includegraphics[width=1\linewidth]{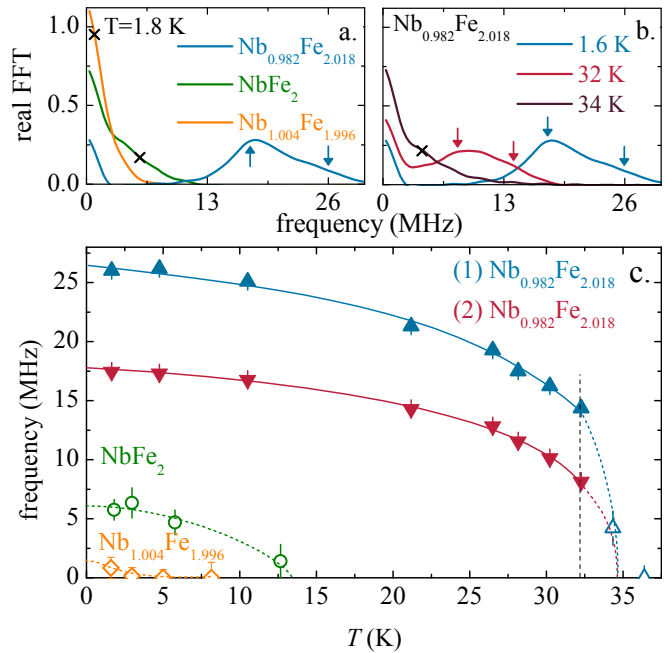}
     \caption{(Color online) (a) Real part of the amplitude of the Fast Fourier Transformation (FFT) of $P_{\mathrm{ZF}}$ of Nb$_{1-y}$Fe$_{2+y}$ at 1.8\,K and (b) for Nb$_{0.982}$Fe$_{2.018}$ at temperatures T $\leq 34$\,K. Here, the arrows ($\Omega_i$ (FM)) and crosses ($\Omega_B$ (SDW)) indicate the characteristic frequencies, see Eq.~\ref{eq:iFld} and \ref{eq:iBsl}.  (c) Temperature dependence of $\mu$SR frequencies of Nb$_{1-y}$Fe$_{2+y}$, $y=+0.018$ (triangle), $0.0$ (circle) and $-0.004$ (diamond). The closed symbols show the frequencies of the FM state, the open one of the SDW state; solid and dashed lines are guides to the eye.}
	\label{fig:frequency}
\end{figure}

Ferromagnetic order is known from bulk magnetometry to set in below about 32\,K in Fe-rich Nb$_{0.982}$Fe$_{2.018}$. Indeed, a spontaneous muon rotation signal is detected at temperatures below 32\,K, consistent with ferromagnetism (Fig.~\ref{fig:ZFlowT}a). The real part of the power spectrum of the low temperature muon relaxation data exhibits a broad maximum at finite frequency (Fig.~\ref{fig:frequency}a), which can be analyzed in terms of two distinct muon sites with precession frequencies $\Omega_i$ caused by the local internal field: 
\begin{equation}
P_{\mathrm{FM}}=\sum^{2}_{i=1}a_i \left[ \alpha_i\cos\left(\Omega_it\right)e^{-\lambda_{\mathrm{T},i}t}+\left(1-\alpha_i\right)e^{-\lambda_{\mathrm{L},i}t}\right].
\label{eq:iFld}
\end{equation}
The asymmetry parameter $a_{i}$ is associated with site $i$, $\alpha_i$ denotes the fraction of transverse field components of the field distribution with respect to the initial muon spin which give rise to a precession. It remains nearly constant as a function of temperature indicating that also the direction of Fe moments stays the same. $\lambda_{\mathrm{T/L}}$ is the transverse/longitudinal damping rate. The frequencies $\Omega_i$ are indicated by arrows in the Figs.~\ref{fig:frequency}a/b, and fits to Eq.~\ref{eq:iFld} as solid lines in Fig.~\ref{fig:ZFlowT}a. The temperature dependence of the precession frequencies $\Omega_i$ (Fig.~\ref{fig:frequency}c) reflects the evolution of the bulk magnetization. 

\begin{figure}[t]
	\centering
		\includegraphics[width=0.9\columnwidth]{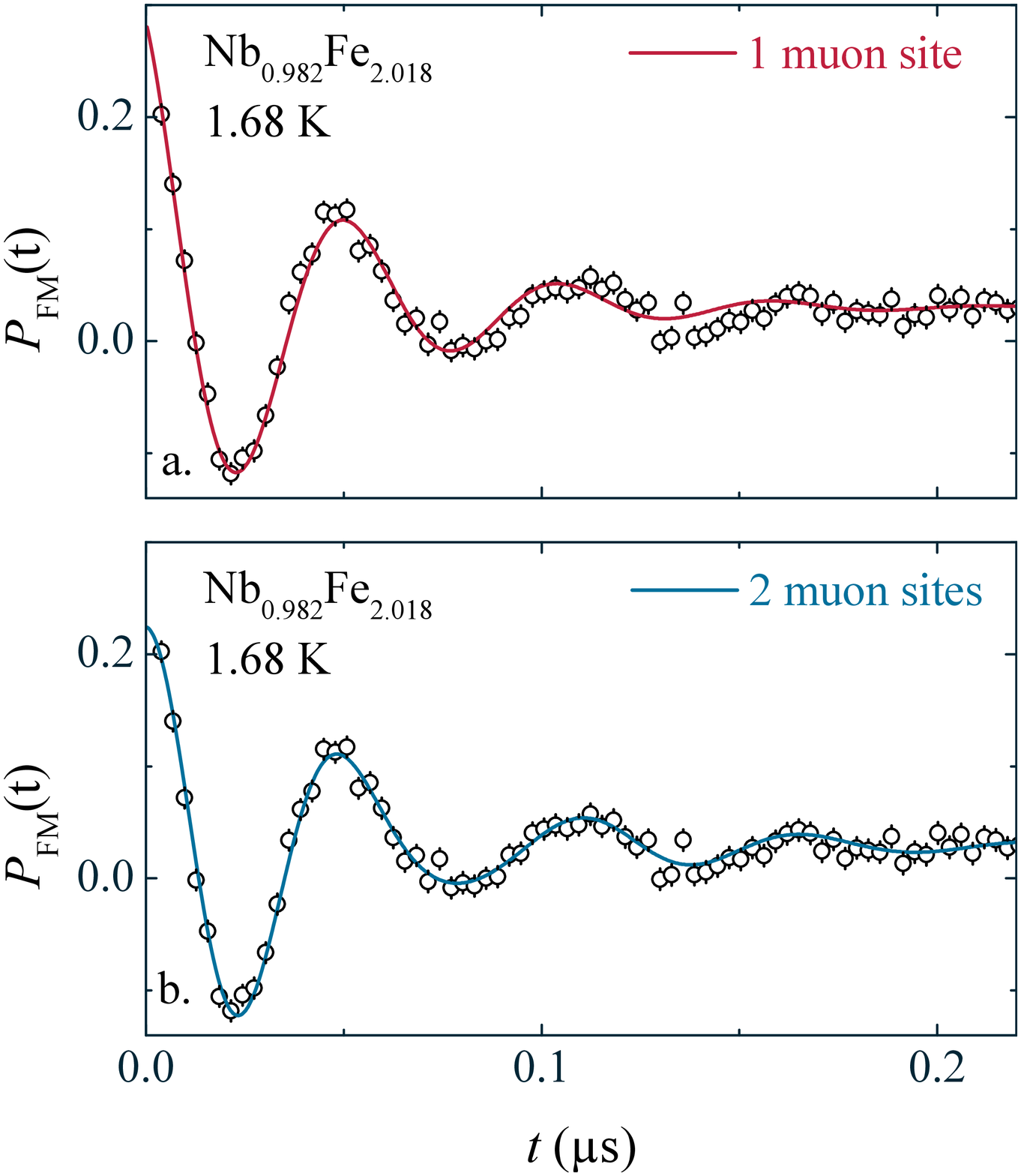}
		\caption{(Color online) Zero field $\mu$SR asymmetry spectrum $P_{\mathrm{FM}}$ of Nb$_{0.982}$Fe$_{2.018}$ at lowest experimental temperature compared to fits of the data assuming muon precession on one and two distinct muon sites, respectively.}
	\label{fig:muonsites1}
\end{figure}

We note that a fit using only one muon site/precession frequency does not properly reproduce the experimental data. This is illustrated in Fig. \ref{fig:muonsites1}, where we plot the muon relaxation spectrum of Fe-rich Nb$_{0.982}$Fe$_{2.018}$ deep in the ferromagnetic phase, together with fits to the data assuming one (red solid line) and two (blue solid line) muon sites, respectively. From the figure, it is evident that a fit using only one muon site fails to reproduce the experimental data already at $\sim 0.05\, \mu$s. Correspondingly, in our fits we have chosen the minimum number of two muon sites to parameterize the experimental data.

\section{Results: M\"{o}ssbauer spectroscopy}

\begin{figure}[t!]
	\centering
		\includegraphics[width=1\linewidth]{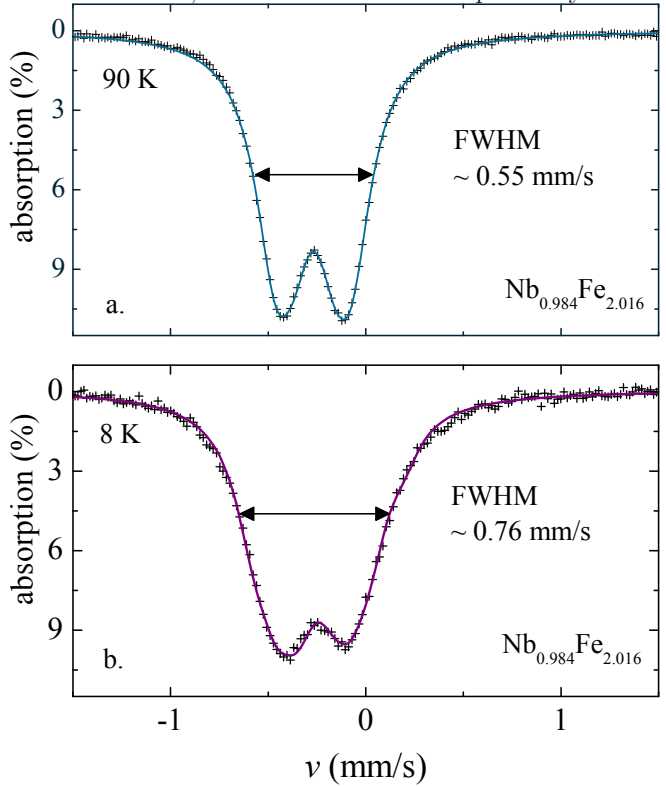}
	\caption{(Color online) M\"{o}ssbauer spectra of Nb$_{0.984}$Fe$_{2.016}$ at (a) 90\,K and (b) 8\,K; the full lines describe the result of a fit to the data, for details see text.}
	\label{fig:moess}
\end{figure}

A M\"{o}ssbauer study on Fe--rich powder Nb$_{0.984}$Fe$_{2.016}$ (T$_C \sim 22$\,K, T$_N \sim 30$\,K from magnetization) yields complementary information on the nature of the magnetic phase below T$_C$. In Fig.~\ref{fig:moess}a/b we plot the M\"ossbauer spectra taken in the paramagnetic (90\,K) and the low temperature magnetic (8\,K) phase \cite{moess14}. At high temperatures the resonance pattern reveals a doublet structure typical for nuclear electric quadrupole interaction. For the fit we take into account two sites, Fe {\it 6h} and Fe {\it 2a} (Fig.~\ref{fig:moess}a), with electric field gradients of about $1.0 \cdot 10^{21}$\,V/m$^2$. Upon lowering the temperature below T$_C$ the onset of magnetic hyperfine splitting at the Fe sites becomes visible from a broadening of the resonance pattern (Fig.~\ref{fig:moess}b). From the full width half maximum FWHM, the magnetic hyperfine field is estimated to be less than 2\,$T$, which corresponds to an upper limit of the ordered magnetic moment of about $0.15\,\mu_B /{\rm Fe\,atom}$ from scaling to metallic Fe and is consistent with the net moment of $0.06 ~\mu_B/{\rm Fe}$ inferred from bulk magnetometry \cite{Friedemann13}.

These findings contradict the suggestion based on Compton scattering at elevated magnetic fields \cite{haynes12} that Fe-rich Nb$_{0.985}$Fe$_{2.015}$ is ferrimagnetic, with moments of 0.4\,$\mu_B$ (-0.6\,$\mu_B$) on the Fe {\it 6h} (Fe {\it 2a}) site. The resolution of this discrepancy requires further investigation, but it may be attributed to the larger energy window probed by Compton scattering, which could cause large, slowly fluctuating moments to appear to be static. 

\section{Discussion}

We now turn to the SDW phase derived from bulk studies, which according to the phase diagram can be examined in Fe-rich Nb$_{0.982}$Fe$_{2.018}$ between 32\,K and 36\,K as well as in NbFe$_{2}$ and Nb$_{1.004}$Fe$_{1.996}$ below T$_N$. The associated power spectra (Fig.~\ref{fig:frequency}a) of the $\mu$SR ZF data can be seen for NbFe$_{2}$ and Nb$_{1.004}$Fe$_{1.996}$, with an overdamped response centered very near to zero frequency. Equally, for Nb$_{0.982}$Fe$_{2.018}$ between T$_C \sim 32$\,K and T$_N \sim 36$\,K a similar response is observed (Fig.~\ref{fig:frequency}b). It implies that this behavior represents the common signature of the modulated magnetic phase. Therefore, and in view of the various bulk magnetic studies on Nb$_{1-y}$Fe$_{2+y}$ and our ESR results, we ascribe the overdamped asymmetry signal to a wide field distribution caused by a SDW state, which we parametrize \cite{le93, le97} as 
\begin{eqnarray}
P_{\mathrm{SDW}} = a \left[ \alpha j_0 \left( \Omega_\mathrm{B} t\right) e^{-\Lambda_\mathrm{T} t} + \left( 1 - \alpha \right) e^{-\Lambda_\mathrm{L} t}\right],
\label{eq:iBsl} 
\end{eqnarray}
with $a$\,-\,asymmetry, $\alpha$\,-\,fraction, $j_0$\,-\,zeroth order Bessel function and $\Lambda_{\mathrm{T/L}}$\,-\,transverse/longitudinal damping rate. Here, the argument of the Bessel function contains the maximal frequency $\Omega_\mathrm{B}$ of the SDW frequency distribution (corresponding to the maximal local field). The values $\Omega_\mathrm{B}$ are indicated by crosses in Fig.~\ref{fig:frequency}a/b.

Because the $\mu$SR frequencies $\Omega_i$ and $\Omega_B$ depend on the size of the ordered moment for the FM and SDW state, respectively, their evolution with temperature and sample composition is of particular interest (Fig.~\ref{fig:frequency}c). We find that $\Omega_B({\rm T})$ in NbFe$_2$ is consistent with a second order phase transition at T$_N \simeq 13$\,K, in good agreement with bulk and wTF data. Moreover, $\Omega_B ({\rm T}=0) \simeq 6$\,MHz, which is about twice as large as the value for Nb$_{0.982}$Fe$_{2.018}$ at 35\,K (open triangle in Fig.~\ref{fig:frequency}c). If the SDW amplitude $M_Q$ scales with $\Omega_B$ in the same way as the FM moment $M_0$ scales with $\Omega_i$, then from bulk measurements of $M_0$, which give about $0.06 \mu_B /{\rm Fe}$ in Nb$_{0.985}$Fe$_{2.015}$ (see Ref. \cite{Friedemann13}), we estimate at low temperature the SDW amplitude $M_Q$ to $\sim 0.01 \mu_B /{\rm Fe}$ in NbFe$_{2}$ and about $5\cdot 10^{-3} \mu_B / {\rm Fe}$ in Nb$_{1.004}$Fe$_{1.996}$, which has T$_N \simeq 8$\,K according to bulk magnetometry and $\mu$SR wTF data. This rapid drop of the size of the magnetic moment by nearly an order of magnitude with varying composition $y$, when T$_N$ appears to be reduced only by about a third, may be connected to the very shallow temperature dependence of $\Omega_B$ in Nb$_{1.004}$Fe$_{1.996}$ for 5\,K\,$<$\,T\,$<$\,8\,K. The latter is reminiscent of the case of URu$_2$Si$_2$ \cite{broholm87, matsuda01}. It might be attributed to sample inhomogeneity, which would cause a spread of T$_N$, or to a more intrinsic form of phase separation (see, e.g. \cite{uemura07}). The wTF data (Fig.~\ref{fig:asymmetryTF}a) indicate nearly 100\% magnetic volume fraction at low temperature but further studies with higher temperature resolution are required to check for phase separation in the range  between 5 and 8\,K. 

If sample inhomogeneity can be ruled out, then our $\mu$SR data seem to suggest a rather unusual evolution of the magnetic moment with composition. A naive linear extrapolation of the $\mu_{ord}(y)$-data from our $\mu$SR-study would imply that the magnetic moment is fully suppressed for a composition $y \sim -0.01$, distinctly different from the value $y_c = -0.015$ obtained from bulk studies. This raises the question about the nature of the magnetically ordered phase in the intermediate range $-0.01 \leq y \leq y_c$, and the possibility of the occurrence of a partially ordered phase \cite{uemura07,pfleiderer04} close to the SDW quantum critical point.

In conclusion, our combined microscopic study of the magnetic ground state of Nb$_{1-y}$Fe$_{2+y}$ establishes the bulk nature and spatial homogeneity of the magnetically ordered phases at the lowest temperatures. We have determine the temperature evolution and magnitude of the internal magnetic field, and from this derived the size of the ordered magnetic moments in the different phases. We confirm the FM nature of the low temperature state in Fe-rich NbFe$_2$, and demonstrate the existence of a modulated magnetic phase at low temperature in near-stoichiometric NbFe$_2$. Finally we observe a rather unusual dependence of the ordered moment of the modulated magnetic phase on Nb concentration, which .

We thank Z. H\"usges for helping to orient the crystals and M. Baenitz, C. Geibel, S. Friedemann, G. G. Lonzarich and P. Niklowitz for discussions. We gratefully acknowledge financial support through the German Science Foundation under FOR960 (Quantum Phase Transitions) and from the EPSRC of the UK.

\end{document}